\documentclass[twocolumn,aps,superscriptaddress,floatfix,final,showpacs,prb,eqsecnum]{revtex4}
\usepackage{amsmath}
\usepackage{graphicx}
\usepackage{amssymb}

\makeatletter
\usepackage{epsfig}
\usepackage{dcolumn}
\usepackage{bm}
\usepackage{amsfonts}
\usepackage{graphicx}

\begin{document}

\title{Phase diagram, energy scales and nonlocal correlations in the Anderson lattice model}

\author{D. Tanaskovi\'{c}}

\affiliation{Scientific Computing Laboratory, Institute of Physics
Belgrade, University of Belgrade, Pregrevica 118, 11080 Belgrade,
Serbia}
\author{K. Haule}
\affiliation{Department of Physics and Astronomy, Rutgers
University, Piscataway, New Jersey 08854, USA}
\author{G. Kotliar}
\affiliation{Department of Physics and Astronomy, Rutgers
University, Piscataway, New Jersey 08854, USA}
\author{V. Dobrosavljevi\'{c}}
\affiliation{Department of Physics and National High Magnetic
Field Laboratory, Florida State University, Tallahassee, Florida
32306, USA}

\begin{abstract}

We study the Anderson lattice model with one f-orbital per lattice
site as the simplest model which describes generic features of
heavy fermion materials. The resistivity and magnetic
susceptibility results obtained within dynamical mean field theory
(DMFT) for a nearly half-filled conduction band show the existence 
of a single energy scale $T^*$ which is similar to the single ion
Kondo temperature $T_K^o$.
To determine the importance of inter-site correlations, we have
also solved the model within cellular DMFT (CDMFT) with two sites in
a unit cell. The antiferromagnetic region on the phase diagram is
much narrower than in the single-site solution, having a smaller
critical hybridization $V_c$ and N\'eel temperature $T_N$.
At temperatures above $T_N$ the 
nonlocal correlations are small, and the DMFT paramagnetic solution is 
in this case practically exact, which justifies the ab initio LDA+DMFT approach in
theoretical studies of heavy fermions. Strong inter-site correlations 
in the CDMFT solution for $T<T_N$, however, indicate
that they have to be properly treated in order to unravel 
the physical properties near the quantum critical point.

\end{abstract}

\pacs{71.27.+a,71.30.+h}

\maketitle

\section {Introduction}

Heavy fermions have been intensively studied in the past thirty years
and a large amount of experimental data has been gathered,\cite{StewartRMP1984,Hewson-book} but a
complete microscopic theory of these materials is still not
available.\cite{LohneysenRMP2007} The unusual low temperature
properties originate from the electrons from partially filled
f-shells which hybridize with a broad band of weakly interacting
conduction electrons.
At high temperatures, the f-electrons are weakly coupled to the
conduction electron band and act as local magnetic moments on which the
conduction electrons are scattered. Below the characteristic
temperature (the lattice coherence temperature) $T^*$, the coherent
quasiparticles start to develop and the resistivity suddenly
decreases. The dependence of $T^*$ on microscopic parameters and
the nature of the coherent heavy electron (Kondo) liquid is still a
subject of active
debate.\cite{YangNature2008,NakatsujiTwofluid2004,BurdinPRB2009}

Many experiments clearly show the existence of a unique energy scale 
that characterizes all transport and thermodynamic 
properties\cite{ThompsonPRB1985,McElfreshPRB1990} and 
several attempts were made to explain universal features of heavy fermions,
both within the phenomenological theory\cite{YangNature2008,NakatsujiTwofluid2004} 
and from the solution of the microscopical model.\cite{GrenzebachPRB2006}
There is, however, a growing evidence\cite{BurdinPRB2009} that 
the energy scales which dominate the low temperature properties of 
heavy fermions depend on details of the
density of states near the Fermi level and the degeneracy and 
crystal fields splitting of the f states - the system dependent 
properties which cannot be captured by the simple theoretical model 
with just one f-spin dublet or a featureless conduction band density 
of states. The physics is even richer at temperatures
$T\ll T^*$, where the system typically orders magnetically and even 
exhibits superconductivity.\cite{GegenwartNatPhys2008,HeggerPRL2000,PetrovicJPCM2001}

In this work we solve the Anderson lattice model (ALM) with
one f-electron orbital per lattice site, in order to precisely 
determine the lattice coherence temperature $T^*$ and 
the importance of nonlocal correlations in different regions of the phase
diagram.
We concentrate on the most interesting regime of
parameters near the antiferromagnetic phase driven by the
conduction electron mediated Ruderman-Kittel-Kasuya-Yosida (RKKY)
interaction.\cite{TahvildarPRB1997} The model is first solved within the DMFT 
approximation\cite{Georges1996} which
is exact in the case of purely local correlations, i.e.~in the
case where the self-energy depends only on frequency and not on
the momentum. The relevance of the local approximation is tested
by a comparison with the CDMFT solution.\cite{KotliarCDMFT2001,JarrellRMP2005} 
We consider a cluster of two sites in a self-consistently
determined medium as a minimal model which treats the inter-site
correlations beyond the mean field level. For temperatures larger
than the N\'eel temperature, we find that the nonlocal
correlations are very small and the local DMFT solutions becomes
practically exact. Therefore, for stronger hybridization the
lattice coherence temperature is determined by the local DMFT
solution and in this case, in the ALM close to half-filling, 
we find that it is proportional to the
single ion Kondo temperature for the same set of parameters, 
$T^* \approx T^*_{DMFT} \sim T^o_K$.
For weaker hybridization, near the antiferromagnetic critical
point, $ T^*_{DMFT}  \lesssim T_N$ and the coherence temperature
is likely to be dominated by the inter-site correlations driven 
by the RKKY interaction. To determine the precise form of $T^*$ 
in this regime and to unravel the physical properties near the 
quantum critical point, we need to
consider larger clusters and different clustering schemes. The
important conclusion can, however, be drawn already from the
present results: for temperatures $T > T_N$ the correlations are
local, which means that LDA+DMFT theory gives an excellent
framework for a quantitative study of heavy fermion materials in
this temperature range.\cite{ShimScience2007,Choipreprint2011,Streltsovpreprint2011}
The LDA+DMFT method,\cite{KotliarRMP2006} obtained by combining DMFT
with the local density approximation (LDA) treats on equal footing the
band structure, the atomic multiplet splitting and  the Kondo
physics, but assumes that the correlations are local in space.
This method has led to a significant progress in the study of
strongly correlated materials, and may also prove crucial in order
to determine the importance of the crystal field effects and
atomic multiplets for low temperature properties of various heavy
fermions.

The remaining part of the paper is organized as follows. In
Section II we define the Hamiltonian and describe the CDMFT method
of its solution. Section III contains the phase diagram and a
comparison of the results with a single-site DMFT. The coherence
temperature $T^*_{DMFT}$ is determined from the magnetic
susceptibility and resistivity results in Section IV, and in
Section V the strength of nonlocal correlations is examined.
Conclusions and discussion are presented in Section VI.

\section{Methods}

We consider the periodic Anderson model of three-dimensional cubic
lattice given by the Hamiltonian

\begin{eqnarray} H &=& -t\sum_{\langle ij \rangle ,\sigma}c^{\dag}_{i\sigma}c_{j\sigma} - \mu \sum_{i\sigma}c^{\dag}_{i\sigma}c_{i\sigma} + V \sum_{i\sigma} (f^{\dag}_{i\sigma}c_{i\sigma} + \mbox{h.c.} ) \nonumber \\
&+& (E_f - \mu) \sum_{i\sigma} f^{\dag}_{i\sigma}f_{i\sigma}
+U\sum_{i}n^{f}_{i \uparrow}n^{f}_{i\downarrow} . \label{M1}
\end{eqnarray}
$c_{i \sigma}^{\dag}$ and $f_{i \sigma}^{\dag}$ create a
conduction band electron (c-electron) and f-electron at site $i$ for spin
$\sigma$.  $n^{f}_{i \sigma}=f_{i \sigma}^{\dag}f_{i \sigma}$ is
the occupation number operator of f-electrons, $t$ nearest neighbor hopping
amplitude, $\mu$ chemical potential, $V$ hybridization strength,
$U$ interaction, and $E_f$ is f-electron energy level.
In DMFT, the solution of the ALM reduces to solving a single impurity
problem supplemented by a self-consistency condition.\cite{Georges1996}
In CDMFT, in contrast, the original lattice is tiled with a superlattice of
clusters. An effective Anderson impurity action is derived for a
single cluster and supplemented by the self-consistency condition
which relates the cluster Green's function to the local Green's
function of the superlattice.\cite{KotliarCDMFT2001,JarrellRMP2005}
For the cluster of two impurities, allowing for the
antiferromagnetic order, there are three independent components of
the cluster Green function, e.g.~$G_{11\uparrow}$,
$G_{22\uparrow}$, and $G_{12\uparrow}$. Details of the self-consistent
procedure for calculation of Green's function are presented in Appendix A.

Technically the most difficult step in the DMFT (CDMFT) procedure is a
solution of the model of an impurity (cluster of impurities)
immersed into the given conduction bath. For this step we use the
Continuous Time Quantum Monte Carlo (CTQMC) impurity
solver\cite{Werner2006} in the implementation from
Ref.~\onlinecite{Haule2007}. This allows us to obtain numerically
exact solution even at very low temperatures which are well below
the N\'eel temperature of the model. We note that the same model
in the CDMFT framework was studied previously, but this work used numerical
methods which are inferior as compared to the CTQMC. The CTQMC allows
us to reach temperatures order of magnitude lower than the
Hirsh-Fye impurity solver used in Ref.~\onlinecite{Sun2005}. The
exact diagonalization method,\cite{DeLeo2007} on the other hand,
is restricted to zero temperature, it discretizes the degrees of
freedom of the conduction bath and uses a discrete mesh of
frequency points much larger than the temperature in our work.
Since the energy scales for the range of parameters where the
Kondo temperature and RKKY interaction energy are comparable in magnitude
are very small, the numerical method that we use in this paper is
crucial in order to precisely determine the phase diagram and to
examine the importance of nonlocal correlations.

\section{Phase diagram}

We present the solution of DMFT (CDMFT) equations for the Anderson
lattice model for $U=1.2$, $E_f=-0.4$, $\mu=-0.03$ and various
hybridization $V$. These parameters correspond to metallic nearly
half-filled system, where stable magnetic phase and strong nonlocal 
effects are expected. The occupation of f-electrons is close to $1$
(Kondo limit) and the total occupation close to 2. Nearly
half-filled conduction band leads to  antiferromagnetic
correlations in the spin density. We will concentrate on the most
interesting regime of hybridization where the Kondo temperature and
RKKY interaction energy are of the same
order of magnitude. The energy will be measured in units of the
conduction electron half-bandwidth $D=6t=1$. The lowest
temperature in numerical results is $T=1/1200$ which is crucial in
order to stabilize the antiferromagnetic solution within CDMFT.

\begin{figure}[t]
\begin{center}
\includegraphics[  width=3. in,
keepaspectratio]{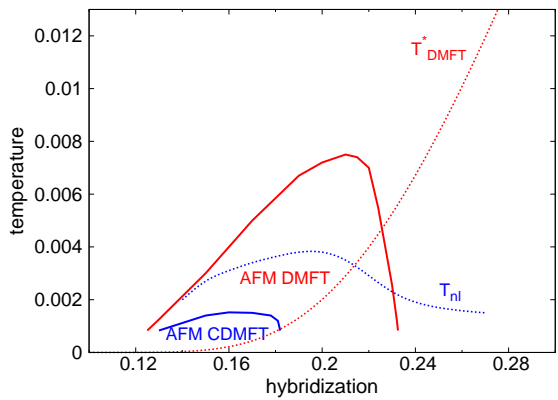}
\caption{ (Color online) Temperature vs.~hybridization phase
diagram in DMFT (solid red line) and CDMFT (solid blue line). The red
dotted line is the coherence temperature in the DMFT solution.
Nonlocal correlations are very weak for temperatures above
$T_{nl}$ (blue dotted line) and in this region the paramagnetic
DMFT solution is practically exact.}
\label{phasediagram}
\end{center}
\end{figure}

The phase diagram of the model is shown in Fig.~\ref{phasediagram}. 
The phase boundary between the
antiferromagnetic (AFM) and paramagnetic solution is determined by
the relative strength of the Kondo screening and RKKY
interaction. The result is in a qualitative agreement with
Doniach's phase diagram:
the AFM solution is stabilized for small
hybridization $V$ when $J_{RKKY} \sim \rho_o {J_K^o}^2$ dominates
over the Kondo scale $T_K^o \sim \exp(-1/2\rho_o J_K^o)$. Here
$J_K^o=(\frac{1}{|E_f-\mu|} +\frac{1}{|U+E_f-\mu|})V^2$ is the
bare Kondo coupling and $\rho_o$ is the density of states of the
conduction electrons at the Fermi level. The numerical solution of
ALM model, however, gives us a possibility to {\it quantitatively}
determine the relevant energy scales. Red dotted line in 
Fig.~\ref{phasediagram} is the
lattice coherence temperature $T^*_{DMFT}$ obtained, in DMFT
solution, as the temperature corresponding to the maximum
resistivity for a given value of $V$ (see Section IV). In DMFT,
which neglects nonlocal correlations, the N\'eel temperature
$T_N^{DMFT}$ can be taken as the measure of $J_{RKKY}$. In AFM
phase $T^*_{DMFT}<T_N^{DMFT}$ in almost entire phase diagram (except very
close to the critical $V_c^{DMFT}$), in agreement with recent DMFT phase 
diagram for the Kondo lattice model.\cite{OtsukiPRL2009}

The AFM region in the CDMFT solution is significantly narrower than
in the single site DMFT solution due to the inter-site
correlations which are treated beyond the mean-filed level in
CDMFT. The highest N\'eel temperature in CDMFT is approximately
four times lower than in DMFT.
The critical hybridization $V_c$ for the quantum phase transition
reduces from $V_c^{DMFT}\approx 0.23$ in DMFT to $V_c\approx
0.18$ in CDMFT solution. As examined in detail in Section V, above
the temperature $T_{nl} \sim 0.004$ the nonlocal correlations are
very small and the paramagnetic solution in single-site DMFT
becomes practically exact for $T\gtrsim T_{nl}$. For $T<T_{nl}$,
however, inter-site correlations found in two-site CDMFT are strong
and dominate the low temperature physics of the ALM for $V\lesssim V_c$.

\begin{figure}[t]
\begin{center}
\includegraphics[  width=2.6 in,
keepaspectratio]{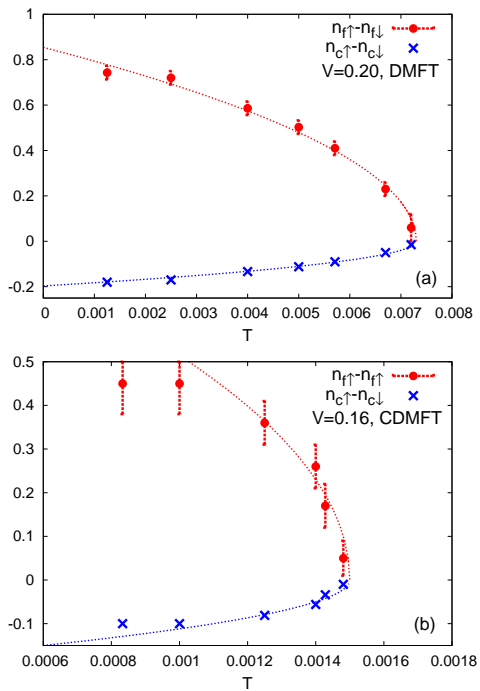}
\caption{ (Color online) Staggered magnetization of f- and
c-electrons in DMFT (a) and CDMFT solution (b). Dotted lines are
fit to the square root mean-field curve.}
\label{magnetization}
\end{center}
\end{figure}

The CTQMC impurity solver\cite{Haule2007} enables us to stabilize
the AFM solution in CDMFT at very low temperatures with a small minus
sign problem. The numerical quality of the data can be verified
from the magnetization results shown on Fig.~\ref{magnetization}.
In the DMFT solution, the mean-field behavior of the staggered
magnetization, $n_{f\uparrow}-n_{f\downarrow} \propto
-(n_{c\uparrow}-n_{c\downarrow}) \propto (1-T/T_c)^{1/2}$, is
observed in a wide temperature range. In the CDMFT solution, the
short range correlations are better taken into account and the
mean-field behavior is restricted to narrower temperature region
near $T_c$. The error bars are the statistical errors estimated
from several CTQMC runs. They are much larger in the CDMFT solution
due to the appearance of a minus sign problem in the AFM phase. The
staggered magnetization $m_f=n_{f\uparrow}-n_{f\downarrow}$ is
less than 1 even as $T\rightarrow 0$ due to the hybridization with
the conduction electrons.  $m_f$ in CDMFT solution is almost two
times smaller than in DMFT. Typical results for the self-energy 
and Green's functions on the Matsubara axis are shown in Appendix B.

\section{Coherence temperature in DMFT solution}

At high temperatures f-electrons are weakly coupled to the
conduction band electrons and behave as local moments. The
scattering of c-electrons initially increases with decreasing
temperature similarly as in the limit of diluted magnetic moments.
The resistivity reaches a maximum at a characteristic temperature
$T_{max}$ that can be taken for a definition of the lattice
coherence temperature. Below $T_{max}$ f- and c-electrons strongly
hybridize and eventually form long-lived heavy
quasiparticles.

In the single-site DMFT it is easy to calculate the scattering rate
and the resistivity. They are obtained from the self-energy $\Sigma_c$
which corresponds to the conduction electrons. The conduction
electrons Green function is given by $ G_c(\omega)=\frac{1}{N}\sum_{\vec k} [
\omega+\mu-\varepsilon_{\vec k}-\Sigma_c(\omega) ] ^{-1}, $ where
$ \Sigma_c(\omega)=V^2/(\omega-E_f+\mu-\Sigma_f(\omega)), $ and
$\Sigma_f$ is the self-energy of the impurity (i.e.~f-electron).
The scattering rate is given by $\tau^{-1}=-2\mbox{Im}
\Sigma_c(\omega=0)$, and the resistivity $\rho$ is obtained from
the zero frequency limit of the real part of the optical
conductivity,\cite{SchweitzerPRL1991,MirandaJPCM1996} $\rho = 1/ \mbox{Re}
\, \sigma(\omega \rightarrow 0 ) $,
\begin{equation}\label{rho_dc}
\rho^{-1}=\pi e^2 \frac{1}{N} \sum_{\vec k}\int d \omega \left(
-\frac{df}{d\omega} \right) v_x^2 A^2({\vec k},\omega).
\end{equation}
Here $A({\vec k},\omega)=\mbox{Im}(\omega+\mu-\varepsilon_{\vec
k}-\Sigma_c(\omega)) $ is the conduction electron spectral
function, $v_x=\partial{\varepsilon_{\vec k}}/ \partial k_x$, 
$N$ is the number of $\vec k$ states in the Brillouin zone,
and $f$ is the Fermi-Dirac distribution.
In the CTQMC impurity solver the self-energy is obtained at Matsubara
frequencies and to obtain the real frequency data we assume the
polynomial form for $\Sigma_c$ at low frequencies, $\Sigma_c =
az^2 + bz +c$, and determine the complex parameters $a,b$, and $c$
from the real and imaginary parts of $\Sigma_c(i\omega_n)$ for
first three Matsubara frequencies, for each $T$ and $V$. This 
simple analytical continuation is not restricted to the Fermi
liquid region and it turned out to be remarkably accurate as we
will see from the analysis of the resistivity curves.

\begin{figure}[t]
\begin{center}
\includegraphics[  width=2.6 in,
keepaspectratio]{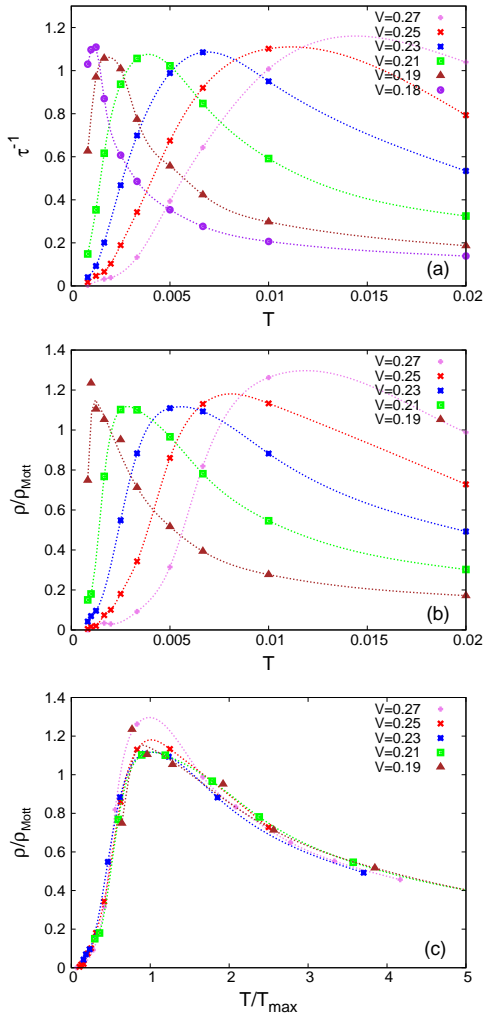}
\caption{ (Color online) (a) Scattering rate and (b) resistivity as a function of
temperature for several hybridization strengths. (c) The
resistivity curves approximately collapse to a single one after
scaling the temperature.}
\label{rho}
\end{center}
\end{figure}

The scattering rate is shown on Fig.~\ref{rho}(a) as a function of
temperature and for several values of hybridization parameter. The
scattering rate curves have a prominent maxima at values which
are of the order of the Mott-Ioffe-Regel limit for maximal metallic
resistivity, $\tau^{-1}_{max}\sim 1$. The resistivity saturation
at this value, which corresponds to the mean free path of one
lattice spacing, is indeed the property of heavy
fermions.\cite{Gunnarsson2003,Hussey2004} It can be simply
explained from the sum rule, and the resistivity saturates when
the Drude peak in the optical conductivity gets completely smeared
with increasing temperature.\cite{NOTE-Mott_limit} The resistivity
curves, Fig.~\ref{rho}(b), have the same form as the scattering
rate curves, with only slightly shifted maxima due to the
temperature dependence of the real part of $\Sigma_c$. The
resistivity is given in units of $\rho_{_{Mott}}$ defined as the
resistivity for $\tau^{-1}=1$. When the temperature is scaled with
$T_{max}$, the shape of the resistivity curves is almost the same
for all values of the hybridization strength, Fig.~\ref{rho}(c).

\begin{figure}[t]
\begin{center}
\includegraphics[  width=3. in,
keepaspectratio]{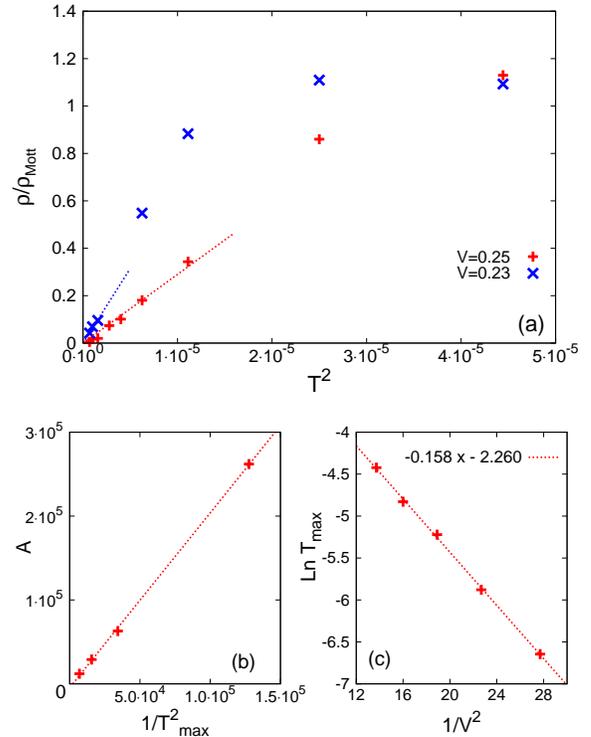} \caption{ (Color online) (a) Resistivity as
a function of $T^2$. The linear region is observed for $T\lesssim
T_{max}/2$. (b) The resistivity slope $A$ is a linear function of
$1/T_{max}^2$. (c) The temperature $T_{max}$ of the resistivity
maximum depends exponentially on the hybridization. }
\label{scaling}%
\end{center}
\end{figure}

The resistivity, Fig.~\ref{scaling}(a), follows the Fermi liquid form,
$\rho=AT^2$, up to the temperature $\sim T_{max}/2$ which can be
taken as the boundary of the Fermi liquid region.  We use the data
for $T<T_{max}/2$ to determine the slope $A$, which depends
linearly on $1/T_{max}^2$, Fig.~\ref{scaling}(b). This is a
manifestation of the Kadowaki-Woods 
relation\cite{PowellNatPhys2008,McElfreshPRB1990,KadowakiSSC1986} 
which establishes a universal
ratio between the resistivity and thermodynamic quantities,
such as the specific heat. In our case $A \sim 1/T_{max}^2 \sim
{m^*}^2 \sim \gamma^2$, where $m^*$ is the effective mass and
$\gamma$ is the specific heat coefficient.
The Kadowaki-Woods ratio explains excellent scaling of the
resistivity curves at low temperatures. By scaling only the
temperature, we find that the curves approximately collapse to a
single curve in the whole temperature range since the maximum
resistivity is approximately the same for all values of
hybridization. The resistivity scaling was successfully applied 
in an early experimental paper on $\mbox{CeCu}_6$.\cite{ThompsonPRB1985}

The resistivity scaling clearly shows the existence of just one
energy scale - lattice coherence temperature $T^* \equiv T_{max}$.
Therefore, it is very important to determine its dependence on
microscopic parameters and make a comparison with the single ion
Kondo temperature. As we show on Fig.~\ref{scaling}(c), the
resistivity maximum depends exponentially on the hybridization
parameter. We can use a relation $T^* = C \exp (-1/2\rho_o
J_K^{latt}) $ as a definition for the {\it lattice Kondo coupling}.
Taking $\rho_o=0.855$ for the conduction band density of states, we
obtain $J_K^{latt} = 3.7 V^2 \approx J_K^o$, where $J_K^o=(\frac{1}{|E_f-\mu|} +\frac{1}{|U+E_f-\mu|})V^2= 3.9 V^2$. Therefore, in the
theory with only local correlations, the coherence temperature has
the same functional form as the single ion Kondo temperature
$T_K^o$ and the effective Kondo coupling $J_K^{latt}$ is
approximately the same as  $J_K^o$. We note that the functional
form of $T_{max}(V)$ is the same if $T_{max}$ is taken from the
scattering rate curves, with the same value for $J_K^{latt}$ and with
the prefactor $C$ only slightly smaller than the one obtained from
the resistivity curves.

\begin{figure}[t]
\begin{center}
\includegraphics[  width=2.8 in,
keepaspectratio]{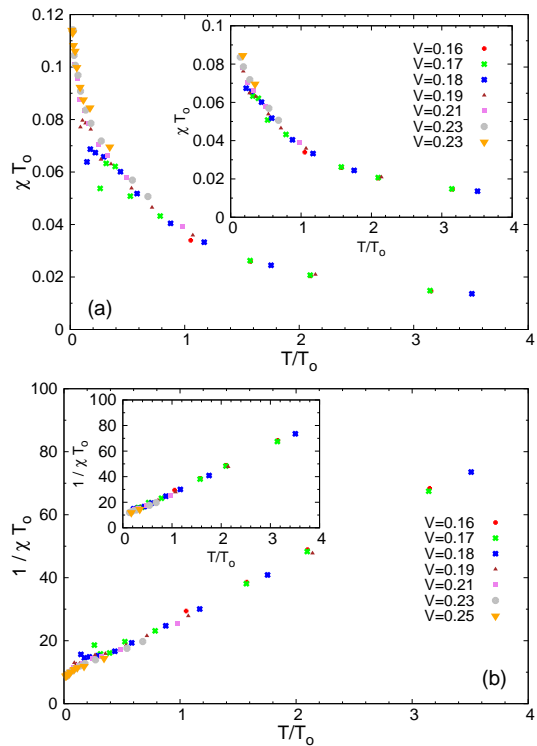}
\caption{ (Color online) Scaled susceptibility (a) (inverse susceptibility (b)) as a function of scaled temperature. If we omit the data for $T<T_{max}$ (the insets), the scaling is excellent.}
\label{susceptibility_scaled}%
\end{center}
\end{figure}

We further compare the lattice and single ion energy scale using
the magnetic susceptibility data. Static local magnetic
susceptibility, $\chi_{loc}(\omega=0) \equiv \chi$,  can be
determined very accurately using CTQMC as the impurity solver and
does not require analytical continuation of the data.
The plots on Fig.~\ref{susceptibility_scaled} are obtained by scaling with
a single parameter $T_o$ - which we call {\it the lattice 
Kondo temperature}. As in the 
single ion case, the temperature is scaled by $T_o$ and the
susceptibility is multiplied by $T_o$ in order to collapse the data on 
a single curve. $T_o$ has exponential dependence on $V^2$ as 
we analyze in detail in the rest of this Section. 
The scaling of the susceptibility,
Fig.~\ref{susceptibility_scaled}(a), is very
good except for the temperatures $T<T_{max}$. The reason is that the
hybridization bath assumes strong temperature dependence for
temperatures lower than the lattice coherence temperature. 
If we omit the data in the scaling analysis for $T<T_{max}$ 
for each value of $V$, 
we find that all the data collapse to a single universal 
curve, inset in Fig.~\ref{susceptibility_scaled}(a). The same scaling
analysis for the inverse susceptibility is shown 
in Fig.~\ref{susceptibility_scaled}(b).

\begin{figure}[t]
\begin{center}
\includegraphics[  width=2.7 in,
keepaspectratio]{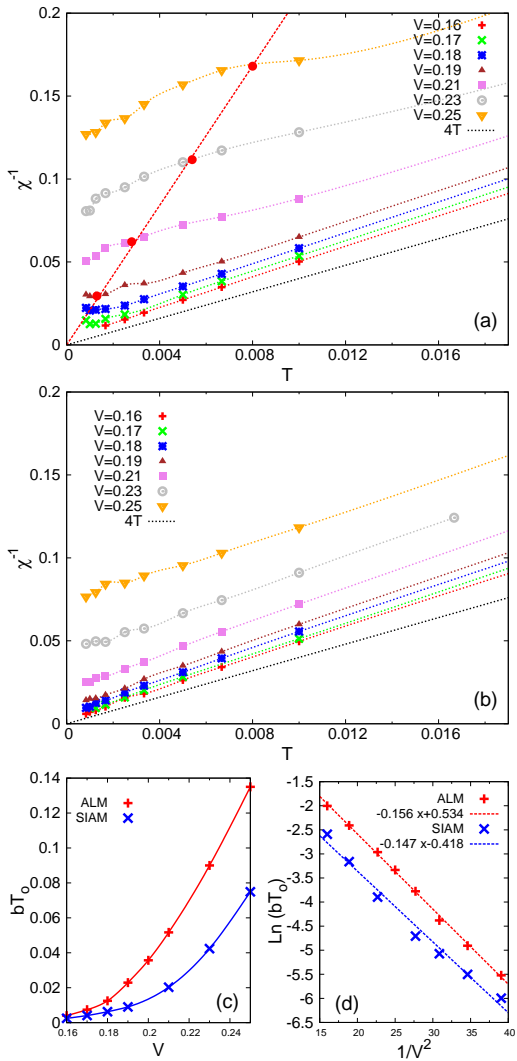}
\caption{ (Color online) Inverse magnetic susceptibility as a function of
temperature for ALM (a), and SIAM (b). For the ALM the
susceptibility follows the Curie-Weiss form above $T^*$ (solid red
dots). The Curie-Weiss temperature for ALM is approximately twice
larger than in SIAM (c), but has the same functional
dependence on hybridization (d).}
\label{inverse_susceptibility}%
\end{center}
\end{figure}

We now carefully analyze the local susceptibility and
make a comparison with the single ion case. 
Fig.~\ref{inverse_susceptibility}(a) and
Fig.~\ref{inverse_susceptibility}(b) show the inverse
susceptibility as a function of temperature for the ALM 
and the single impurity Anderson model (SIAM). 
We start from the Curie-Weiss form
\begin{equation}\label{chi}
\chi^{-1} = aT+bT_o .
\end{equation}
Here $a$ and $b$ are constants. In the single ion case $T_o$
corresponds to the Kondo temperature $T_K^o$. In the lattice model
$\chi(T)$ also follows the Curie-Weiss form except at the
lowest temperatures, $T\lesssim T_{max}$, where it significantly
deviates from linear dependence. The
inverse susceptibility at $T=T^*=T_{max}$ is shown by the solid red dots in
Fig.~\ref{inverse_susceptibility}(a). As expected, the value of
$\chi^{-1}$ at $T=T^*$ is proportional $T^*$. To obtain the
lattice Kondo temperature $T_K^{latt}$, we omit the data for
$T<T^*$ and make a fit to Eq.~(\ref{chi}). For the case of a
single impurity we can keep all data to obtain $T_o \equiv T_K^o$.
The ALM and SIAM values for $T_o$ differ by a factor two,
Fig.~\ref{inverse_susceptibility}(c), but have the same
exponential dependence on $V^2$,
Fig.~\ref{inverse_susceptibility}(d): $T_o \propto
\exp(-1/2\rho_o J_K)$, where the lattice Kondo coupling $J_K^{latt}
\approx J_K^o \propto V^2$. Some deviation from linear behavior
for SIAM is due to the small change of the occupation number 
($0.9<n_f<0.96$) since
we keep the chemical potential fixed while changing $T$ and $V$.

We can conclude that both the resistivity and magnetic
susceptibility data give the same value for the effective lattice
Kondo coupling whose value is very similar to the bare Kondo
coupling in the the case of diluted impurities. 
The Curie-Weiss form, Eq.~(\ref{chi}), gives the value of $T_o$
up to the prefactor. In order to compare the absolute values of
$T^* \equiv T_{max}$ and $T_K^{latt} \equiv T_o$ for the ALM, we 
can use the value $b$ from the
single impurity theory. From the Wilson formula\cite{Hewson-book}
$\chi(T=0) = 0.4128/(4T_K^o)$, which gives $b=9.7$. 
(We used $b=9.7$, $a=4$, and $ g \mu_B=1$ on the scaling 
plot, Fig.~\ref{susceptibility_scaled}.)
Then $T^*/T_K^{latt}= 0.6$ and their ratio does not depend on $V$.
The conclusion that $T^*$ and $T_K^o$ have the same exponential
dependence on the hybridization parameter, i.e.~on the coupling
constant, agrees with the previous results using the numerical 
renormalization group as the impurity solver\cite{PruschkePRB2000,
GrenzebachPRB2006}, and slave boson mean field theory,\cite{BurdinPRB2009}
and this is valid even far away from half-filling. 
The prefactor is, however, of the order of 1 
only in the case of nearly half-filled featureless conduction band.
Only in this case there is a single energy scale in the ALM and all
additional low-energy scales assigned to different physical
properties are proportional to this single low-temperature scale.
For small occupation of c-electrons there are
two energy scales: $T_K^0$ where the screening begins, and 
$T^* \ll T_K^o$ where coherence sets in.\cite{PruschkePRB2000,
GrenzebachPRB2006} $T^* \ll T_K^o$ also if there is a peak 
in the noninteracting conduction band density of states, while
$T^* \gg T_K^o$ if there is a dip at the Fermi level.\cite{BurdinPRB2009}
Before we concentrate on the strength of inter-site
correlations, which has not been previously explored, we will make few
additional remarks about the analytical continuation performed in our work.

The only assumption that we use is that the self-energy is an analytical
function, and we approximate the low-frequency part by a second order
polynomial obtained from the self-energy at first three Matsubara frequencies.
High frequency part of the self-energy is not important at
all when calculating the resistivity, since the derivative of the Fermi-Dirac
function in Eq.~(\ref{rho_dc}) is negligible away from the Fermi level. It is enough to
keep the frequencies $|\omega|\lesssim 3T$ in the integral, and in this case
a second order polynomial is a reasonable approximation for the self-energy.
The approximation by a polynomial would be problematic if the self-energy is
non-analytic near the quantum critical point. However, in our case we do not have
such an irregular self-energy to worry about. Finally, our results
include two stringent tests of the analytical continuation: the Fermi
liquid behavior at low $T$ is reproduced remarkably well, and the susceptibility
data, which does not require the analytical continuation, give the same energy
scales as obtained from the resistivity calculations. We emphasize, however,
that our method for analytical continuation is not restricted to the Fermi
liquid region, and we believe that it is the best possible option if we
are interested only in the low frequency part of the spectrum. The maximum
entropy method gives roughly correct spectra at intermediate frequencies,
but from our experience, it never gives better results than the polynomial
fit at low frequencies. The small noise from QMC data can also lead to fairly
bad results in the Pad\'e method for the analytical continuation.

\section{Strength of nonlocal correlations}

\begin{figure}[t]
\begin{center}
\includegraphics[  width=2.5 in,
keepaspectratio]{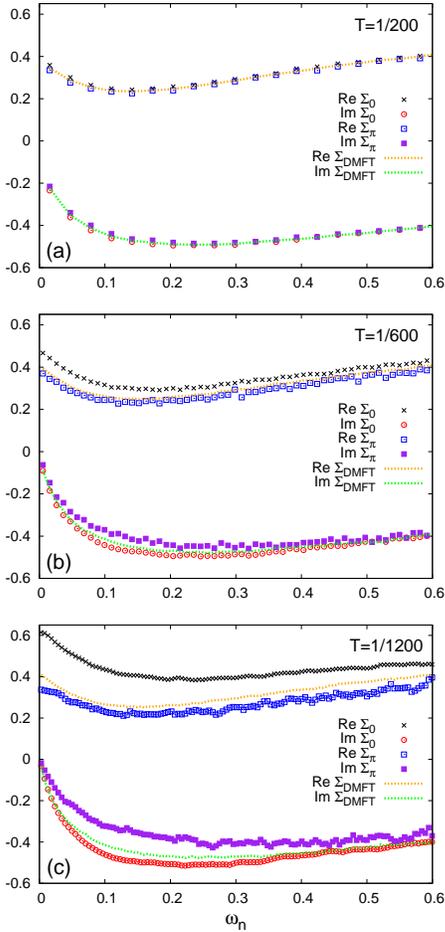}
\caption{ (Color online) Comparison of the paramagnetic DMFT and CDMFT
solution for the self-energy on the Matsubara axis for 
$V=0.18$ and $T=1/200, 1/600, 1/1200$.}
\label{nonlocality}%
\end{center}
\end{figure}

The results obtained in the previous section are exact if the
correlations are local, i.e.~if the self-energy depends only on
the frequency and not on the momentum. In order to determine the
importance of nonlocal correlations we consider the ALM within
CDMFT with two sites in a unit cell, as the minimal model which
includes nonlocal correlations. We restrict to the paramagnetic
solution. Typical results for the f-electron self-energy are shown
in Fig.~\ref{nonlocality}. The hybridization parameter in this
figure is chosen very close to critical value $V=0.18\approx V_c$, but the
results for the self-energy are qualitatively the same also for
hybridization away from the critical point. The inter-site
correlations are determined by the difference between even and odd
components of the self energy, $\Sigma_{00}$ and
$\Sigma_{\pi\pi}$. At $T=1/200$, Fig.~\ref{nonlocality}(a),  the
self-energy fully coincides with the single site DMFT solution. At
$T=1/600$, Fig.~\ref{nonlocality}(b), very weak inter-site
correlations are present, and they gradually increase as the
temperature is further lowered to $T=1/1200$, Fig.~\ref{nonlocality}(c). 
We note that we did not find any signatures of the Kondo 
breakdown - the decoupling of f-electrons and Fermi surface
reconstruction for $V=V_c$.\cite{SenthilPRB2004} The imaginary part 
of the self-energy $\mbox{Im} \Sigma_f (\omega = 0)$ goes to zero, 
and the quasiparticle weight, 
$Z=(1-\partial \mbox {Im} \Sigma(i\omega)/ \partial \omega )
^{-1}|_{\omega \rightarrow 0^+} \sim 1/m^*$,
remains finite as $T \rightarrow 0$ and $V=V_c$.

\begin{figure}[t]
\begin{center}
\includegraphics[  width=2.6 in,
keepaspectratio]{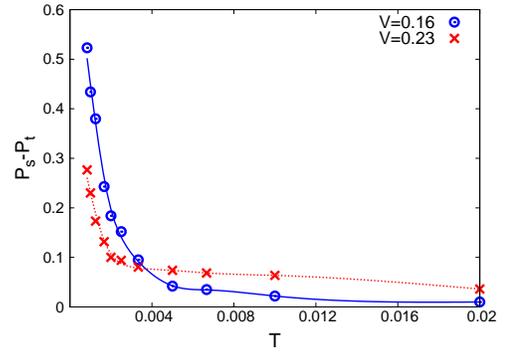}
\caption{ (Color online) Difference of probabilities for finding the two-site impurity in
the singlet and triplet cluster eigenstate. }
\label{probabilities}
\end{center}
\end{figure}

The strength of nonlocal correlations can be quantified using the
probabilities for the occupation of different cluster eigenstates.
At high temperatures, f-electrons are almost decoupled, and the
probabilities $P_s$ and $P_t$ for the singlet and (one of three available) triplet
states are almost the same, and approach to the free spin 
value $P_s\approx P_t \sim 0.25$. At low
temperature the probability of the singlet state suddenly increases [Fig.~8],
and the singlet cluster eigenstate is dominantly occupied for
$T\lesssim T_{nl}$. Large singlet probability $P_s$, which approaches 
to 1 at the lowest temperatures, implies strong singlet correlations. 
We can use these probabilities to define a crossover temperature $T_{nl}$ 
which divides the regions of strong and weak inter-site correlations. 
We define $T_{nl}$ as the temperature when $P_s-P_t=0.1$, which is shown by
a blue dotted line on the phase diagram, Fig.~\ref{phasediagram}. 
$T_{nl}$ roughly follows $T_N$, but $P_s-P_t$ stays large at the lowest 
temperatures also for $V>V_c$. The calculation of the observables, 
such as the spin susceptibility and resistivity, remains to be 
done in future work. While for a reliable quantitative
analysis we need to study also larger clusters and compare different clustering
schemes since the two-site cluster version may overestimate the local singlet formation, 
we expect that the two impurity results already give a
good estimate of the line which separates the regions of strong and weak inter-site
correlations.

\section{Conclusion and discussion}

In summary, we have solved CDMFT equations with two sites in a
unit cell for a nearly half-filled Anderson lattice model and
compared the results with single-site DMFT. The phase diagram
generally agrees with Doniach's physical picture: the
antiferromagnetic phase is stabilized when RKKY interaction energy is
larger than the Kondo temperature. The CDMFT solution gives much
narrower AFM phase as compared to DMFT, which is expected since
the mean-field solution generally overestimates a tendency to
magnetic order, and two-site CDMFT
overestimates the local singlet-formation, which competes with the long
range magnetic order,  hence  the exact N\'eel temperature of the ALM is
expected to be somewhere between the two limits.
At temperatures above $T_N$ the 
nonlocal correlations are small and the CDMFT and the DMFT paramagnetic solution 
are almost the same. 
This conclusion has  important practical
consequences for theoretical studies of heavy fermions. For
temperatures larger than $T_N$ the self-energy is
weakly momentum-dependent, which explains the
success of local LDA+DMFT approach in ab initio calculation
of transport and thermodynamic properties of heavy fermions. 
Heavy fermions are
particularly well suited for the single-site DMFT approach since the
interesting crossovers in transport and thermodynamic properties,
from coherent to fully incoherent behavior, are seen in a broad
temperature region above very low ordering temperature. At
temperatures $T \lesssim T_N$ when short range processes are
included, and if frustration at short distances is weak, the
modifications from local DMFT predictions can be substantial.

We have also determined the lattice coherence temperature $T^*$ from
the resistivity and magnetic susceptibility calculated within DMFT and
made a careful comparison with the Kondo scale $T_K^o$ for diluted
impurities with the same set of parameters. The results clearly
show that there exists a single energy scale $T^* \sim T_K^o$, which dominates 
the low temperature properties in the case of a nearly 
half-filled featureless conduction band.
The comparison with the CDMFT solution shows that for stronger
hybridization the nonlocal
correlations are negligible at temperatures $T^*(V)$ and that
$T^*$ is approximately the same as given by the local DMFT solution. 
Near the quantum critical point, the inter-site correlations have 
to be properly taken into account to determine the lattice coherence scales.
For this purpose, larger clusters and different clustering
schemes also need to be considered, an important research direction 
which is left for future work. 
In real materials the effects of atomic multiplets and
crystal fields, as well as the existence of sharp peaks or dips in the density
of states at the Fermi level may significantly modify the low temperature 
physics\cite{NevidomskyyPRL2009,BurdinPRB2009} as compared to our simple model.

We note that in this work we have concentrated on a broad
temperature range above the quantum critical point and have not
directly addressed an important and controversial question of the
nature of the quantum critical 
point.\cite{GegenwartNatPhys2008,SenthilPRB2004,KnafoNatPhys2009}
Our two-site CDMFT solution, however, shows that f-electron
density of states remains finite at the Fermi level even very near
the critical point and we did not see signatures of the Kondo
breakdown - decoupling of the f-electrons from the conduction bath
at the Fermi level. This agrees with recent studies of the Kondo lattice model within dynamical cluster approximation,\cite{MartinPRB2010} and numerical renormalization group studies of two-impurity Anderson model.\cite{ZhuPRB2011} Further studies 
in this direction are needed, for different parameter regimes and
larger clusters, facilitated with the CTQMC impurity solver which
is proven to be able to reliably treat the competition of small
energy scales.

\begin{acknowledgments}
We thank M. Ferrero and M. Vojta for usefull discussions. D.T.
acknowledges support from the Serbian Ministry of Education and Science under
project No. ON171017. K.H. was supported by NSF grant DMR-0746395,
G.K. by NSF DMR-0906943, and V. D. by the National High Magnetic
Field Laboratory and the NSF Grant DMR-1005751. D.T was supported
in part by I2CAM under NSF Grant
DMR-0844115. D.T., K.H., and G.K. acknowledge the hospitality of
KITP, Santa Barbara, under NSF Grant PHY05-51164. Numerical
simulations were run on the AEGIS e-Infrastructure, supported in
part by FP7 projects EGI-InSPIRE, PRACE-1IP and HP-SEE.
\end{acknowledgments}

\appendix

\section{Self-consistency equations}

In the CDMFT the original lattice is tiled with a superlattice of
clusters and an effective Anderson impurity action is derived for
a single cluster and supplemented by the self-consistency
condition which relates the cluster Green's function to the local
Green's function of the superlattice. The hybridization bath for
the Anderson impurity action, the cluster Green function and the
cluster self-energy have inter-site components and can be
conveniently represented in the matrix form. For the cluster of
two impurities the Green function takes the form
\begin{equation}\label{4.6}
\hat{G}_f = \left( \begin{array}{cccc} G_{11\downarrow}
& G_{12 \downarrow} & 0 & 0
\\ G_{21 \downarrow} &
G_{22 \downarrow} & 0 & 0 \\
0 & 0 & G_{11\uparrow} & G_{12 \uparrow} \\
0 & 0 &  G_{21 \uparrow} &G_{22 \uparrow}
\end{array}\right) .
\end{equation}

From the CDMFT self-consistency equation, the hybridization
function $\hat \Delta$ is given by
\begin{equation}\label{sc1}
\hat \Delta(i\omega_n)=i\omega_n + \mu - E_{f} -
\hat{\Sigma}_f(i\omega_n)-\hat{G}_f^{-1}(i\omega_n) ,
\end{equation}
where the cluster Green function coincides with the local
component of the lattice Green function
\begin{equation}
\hat{G}_{f}(i\omega_n) = \frac{1}{N}\sum_{\vec k} \hat{G}_{f}(i\omega_n,{\vec
k}) .
\end{equation}
$\hat{G}_{f}(i\omega_n,{\vec k})$ is easily obtained by
integrating out the conduction electrons from the action which
corresponds to the Hamiltonian (\ref{M1}) and its spin $\sigma$
component is explicitly given by
\begin{eqnarray}\label{4.3}
&& \hat{G}_{f\sigma}(i\omega_n,{\vec k})=\left[ \left( \begin{array}{cc} i\omega_n+\mu-E_f & 0 \\
0 & i\omega_n+\mu-E_f \end{array}\right) \right. \nonumber \\
&&- V^2  \left( i\omega_n+\mu-\hat{t}(\vec{k}) \right) ^{-1} - \left .\left( \begin{array}{cc} \Sigma_{11\sigma} & \Sigma_{12 \sigma} \\
\Sigma_{21 \sigma} & \Sigma_{22 \sigma} \end{array}\right)
\right]^{-1} .
\end{eqnarray}
For a hypercubic lattice the summation over $\vec k$ is done in the
reduced Brillouin zone: $k_x \in (-\frac{\pi}{2},\frac{\pi}{2} )$,
$k_y,k_z \in (-\pi,\pi)$, and the hopping term is equal to
\begin{equation}\label{3.2}
\hat{t}(\vec{k})=\left( \begin{array}{cc} 0 & e^{-ik_x}
\varepsilon_{\vec k}
\\ e^{ik_x} \varepsilon_{\vec k} & 0
\end{array} \right) ,
\end{equation}
with $\varepsilon_{\vec k}=-2t (\cos k_x + \cos k_y + \cos k_z) $.

We solve the two-site Anderson impurity problem
using the CTQMC impurity solver as implemented in
Ref.~\onlinecite{Haule2007}. This requires to switch to the
cluster momenta basis functions, which are in the case of two
sites in a cluster given by
\begin{eqnarray}
| \psi_{0,\sigma} \rangle = \left( | \sigma , 0 \rangle +  | 0, \sigma  \rangle \right) / \sqrt 2 \nonumber \\
| \psi_{\pi,\sigma} \rangle = \left( | \sigma , 0 \rangle -  | 0,
\sigma  \rangle \right) / \sqrt 2 \label{..}
\end{eqnarray}
In this alternate basis, the hopping matrix
is equal to
\begin{equation}\label{3.4}
\hat{t}(\vec k)=\varepsilon_{\vec k} \left( \begin{array}{cc}
\cos{k_x} & i \sin{k_x} \\ -i\sin{k_x} & -\cos{k_x}
\end{array}\right) ,
\end{equation}
and the self-consistency equation is given by
\begin{eqnarray}
&& \left( \begin{array}{cc} G_{00\sigma} & G_{0\pi \sigma} \\
G_{\pi 0\sigma} & G_{\pi \pi \sigma}
\end{array}\right)=\frac{1}{N}\sum_k \left[ \left( \begin{array}{cc} i\omega_n+\mu-E_f & 0 \\
0 & i\omega_n+\mu-E_f \end{array}\right) \right. \nonumber \\
&&- V^2  \left( i\omega_n+\mu-\hat{t}(\vec{k}) \right) ^{-1} - \left .\left( \begin{array}{cc} \Sigma_{00\sigma} & \Sigma_{0 \pi \sigma} \\
\Sigma_{\pi 0\sigma} & \Sigma_{\pi \pi\sigma} \end{array}\right)
\right]^{-1} , \label{sc2}
\end{eqnarray}
where
\begin{eqnarray}\label{3.5}
&&\left( i\omega_n+\mu-\hat{t}(\vec{k}) \right) ^{-1} = \frac{1}
{(i\omega_n+\mu)^2-\varepsilon_{\vec k}^2} \nonumber \\
&&\times \left( \begin{array}{cc} i\omega_n+\mu+\varepsilon_{\vec k}
\cos{k_x}
& i\varepsilon_{\vec k} \sin{k_x} \\
-i\varepsilon_{\vec k} \sin{k_x} & i\omega_n+\mu- \varepsilon_{\vec
k} \cos{k_x}
\end{array}\right) .
\end{eqnarray}
The components of the Green functions are related to those in
the direct basis as
\begin{eqnarray}\label{4.4}
G_{00\sigma} &=&
(G_{11\sigma}+G_{22\sigma}+G_{21\sigma}+G_{12\sigma})/2 ,
\nonumber \\
G_{0\pi \sigma} &=&
(G_{11\sigma}-G_{22\sigma}+G_{21\sigma}-G_{12\sigma})/2 ,
\nonumber \\
G_{\pi 0\sigma} &=&
(G_{11\sigma}-G_{22\sigma}-G_{21\sigma}+G_{12\sigma})/2 ,
\nonumber \\
G_{\pi \pi\sigma} &=&
(G_{11\sigma}+G_{22\sigma}-G_{21\sigma}-G_{12\sigma})/2 .
\end{eqnarray}

In the AFM phase $G_{11\uparrow}=G_{22\downarrow}$,
$G_{22\uparrow}=G_{11\downarrow}$,
$G_{12\uparrow}=G_{21\downarrow}$, and
$G_{21\uparrow}=G_{12\downarrow}$. Therefore,
$G_{00\uparrow}=G_{00\downarrow}$, $G_{\pi \pi \uparrow}= G_{\pi
\pi \downarrow}$, $G_{0 \pi \uparrow}= - G_{0 \pi \downarrow}$,
and $G_{\pi 0 \uparrow}= - G_{\pi 0 \downarrow}$. Also, the
off-diagonal Green's functions at constant spin are the same,
$G_{0\pi \uparrow}=G_{\pi 0 \uparrow}$ and $G_{0\pi
\downarrow}=G_{\pi 0 \downarrow}$. Analogous relations are valid
for the self-energy and for the hybridization bath.
Therefore, the effective two-impurity Anderson model is
solved in the hybridization bath with three independent components
\begin{equation}\label{4.5}
\hat{\Delta}=  \left( \begin{array}{cccc} \Delta_{00} &
\Delta_{0\pi} & 0 & 0
\\ \Delta_{0 \pi} &
\Delta_{\pi \pi} & 0 & 0 \\
0 & 0 & \Delta_{00} & -\Delta_{0\pi} \\
0 & 0 & -\Delta_{0\pi} & \Delta_{\pi \pi}
\end{array}\right) ,
\end{equation}
and supplemented by the self-consistency condion, Eqs.~(\ref{sc1})
and (\ref{sc2}).
The Green's function also has three independent components
\begin{eqnarray}\label{4.7}
G_{00}&=& (G_{11}+G_{22})/2+G_{12} ,
\nonumber \\
G_{0\pi}&=& (G_{11}-G_{22})/2 ,
\nonumber \\
G_{\pi \pi}&=& (G_{11}+G_{22})/2-G_{12} ,
\end{eqnarray}
where the spin index has been suppressed. Analogous relations are
valid for the self-energy. We note that the off-diagonal
components $\Delta_{0\pi}$ lead to the antiferromagnetic order. In
the paramagnetic solution they are equal to zero.

\section{Green's functions in the AFM solution}

\begin{figure}[h]
\begin{center}
\includegraphics[  width=2.8 in,
keepaspectratio]{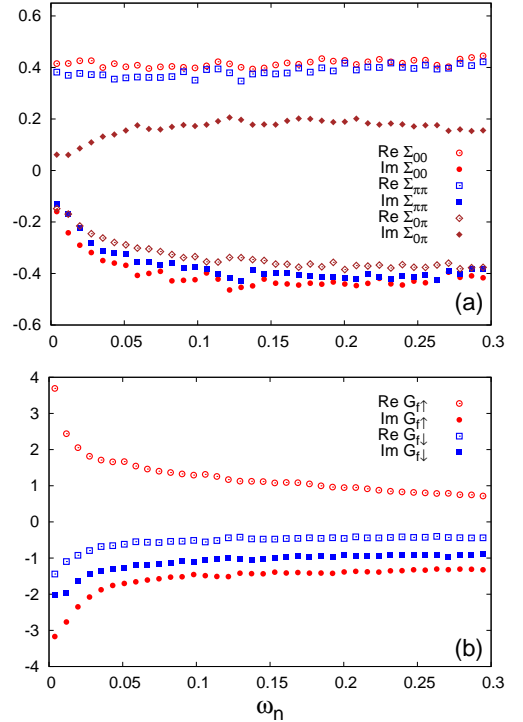}
\caption{ (Color online) Self-energy (a) and local Green's function (b) in CDMFT
solution for $V=0.16$ and $T=1/800$.}
\label{greenfunction}%
\end{center}
\end{figure}

Typical results for the f-electron self-energy and Green's function in the AFM
phase are given in Fig.~\ref{greenfunction}. The self-energy,
Fig.~\ref{greenfunction}(a), has very small nonlocal component
$\Sigma_{12}=(\Sigma_{00}-\Sigma_{\pi\pi})/2$. Finite
$\Sigma_{0\pi}$ component leads to the staggered magnetization.
The corresponding local Green's function,
$G_{11,\sigma}=(G_{00,\sigma}+G_{\pi \pi,\sigma} ) /2 +
G_{0\pi,\sigma}$, has different spin up and spin down components,
Fig.~\ref{greenfunction}(b). For given parameters,
$n_{f\uparrow}-n_{f\downarrow} =0.35$,
$n_{c\uparrow}-n_{c\downarrow} =-0.08$,
$n_{f\uparrow}+n_{f\downarrow} =0.96$, and the total occupation is
1.92.

We note that the N\'eel temperature strongly depends on the occupation
number. It is the highest in the Kondo insulator (for $n_f + n_c =2$),
and drops sharply as the occupation number decreases. In the Kondo
insulator for $V=0.18$, $E_f=-0.6$, we find that $T_N^{DMFT} \approx 0.015$,
which is similar as in Ref.~\onlinecite{TahvildarPRB1997}, while
$T_N^{CDMFT} \approx 0.004$. We suspect that $T_N$ is much larger in
Ref.~\onlinecite{Sun2005} because the solution gets stuck in a metastable
local minimum, giving false higher value for $T_N$, or because of the
self-consistency condition, which is in fact different in Ref.~\onlinecite{Sun2005}.
The expression for the Green function in Ref.~\onlinecite{Sun2005}
includes periodized self-energy which may be noncausal. In our work, we use
the standard CDMFT implementation of the cluster DMFT.

\bibliographystyle{apsrev}

%
\end{document}